\documentstyle[aps,multicol,epsfig]{revtex}

\def \be{\begin{equation}}
\def \ber{\begin{eqnarray}}
\def \ee{\end{equation}}
\def \eer{\end{eqnarray}}
\draft \tolerance=10000

\input epsf

\begin{document}
\title{Imprinted Networks as  Chiral Pumps}
\author{Y.~Mao, M.~Warner}
\address{
Cavendish Laboratory, Madingley Road,\\
Cambridge, CB3 0HE,  UK.}
\date{\today}
\maketitle

\begin{abstract}

We investigate the interaction between a chirally imprinted
network and a solvent of chiral molecules. We find, a liquid
crystalline polymer network is preferentially swollen by one
component of a racemic solvent. This ability to separate is linked 
to the chiral order parameter of the network, and can be reversibly
controlled via temperature or a mechanical deformation. It is maximal
near the point at which the network loses its imprinted structure.
One possible practical application of this effect would be a
mechanical device for sorting mixed chiral molecules.
\end{abstract}

\vspace{0.2cm}
\begin{multicols}{2}

Chirality is of vital importance in pharmacology and chemical
biology. Sorting right handed from left handed molecules is a
common and yet tricky process \cite{chris,israelachvilibook}.
Traditionally, mixtures of chiral molecules are separated using other
chiral molecules with intrinsic chirality \cite{israelachvilibook}
which is often difficult to control. Liquid crystalline
elastomers with an imprinted chirality offers an alternative
solution.

Recently we proposed a theory of chiral imprinting in liquid
crystalline polymer networks \cite{us1} where nematic polymers are
crosslinked in the presence of chiral sol\-vent, that is when they
have an induced cholesteric phase. Subsequent removal of the
solvent, and thus of all the intrinsically chiral material, can
either leave behind a cholesteric network or see the network
lose its chiral structure, depending on the strength of a chiral
order parameter, $\alpha$ (see below). Above a critical point at
$\alpha_c = 2/\pi$, director twist is lost in a second order 
manner. Experiments have shown conclusively that imprinting can be
achieved \cite{Mitchimprint}. 
The mechanical behaviour of such imprinted solids and similarly 
cholesteric nematic networks has been analysed \cite{us2}. 
Imposed strain induces the director to rotate and eventually
eliminates twist. Thus, mechanical deformation can modify and 
even destroy the chiral
structure.  For an imprinted network, this can mean the complete
loss of chirality (in contrast to intrinsic materials that would
thereby simply enter an untwisted chiral nematic phase, N*).
Imprinted networks hence offer chirality that can be controlled
externally. In this article, we calculate the interaction between
the imprinted chiral elastomer and a racemic solvent which is used
to swell the network. We show that the network has the ability to
sort the the components of a mixture according their handedness,
and this ability is a direct result of a chiral order parameter
controllable by simply deforming the sample. This gives rise to
the possibility of  mechanical devices to sort a racemic solvent.

In a cholesteric, the local order is similar to that of a uniaxial
nematic with director $\underline{n}$ (which then takes angles
$\theta_o = q_o x$ with respect to $\underline{z}$ in the $yz$
plane as it advances in a helix). A model for the total free
network energy per chain, $F_n$, in a nematic elastomer is a
simple generalisation\cite{mark} of classical Gaussian rubber
elasticity:
\be
F_n={k_BT \over 2}\;Tr \left[\;\;\underline{\underline{l}}_o\;\;
\underline{\underline{\lambda}}^T \;\;
\underline{\underline{l}}^{-1}\;\;
\underline{\underline{\lambda}}\;\;\right]. \ee The reduced
molecular shape tensors $\underline{\underline{l}}_o$ and
$\underline{\underline{l}}$ before and after deformation record
the director and the intrinsic anisotropy, $r$:
$\underline{\underline{l}}_o =
(r-1)\;\underline{n}_o\;\underline{n}_o
+\underline{\underline{I}}$ and (the inverse) $
\underline{\underline{l}}^{-1} =({1 \over r}-1)
\;\underline{n}\;\underline{n} +\underline{\underline{I}}$.

As a result of swelling, we expect the deformation tensor
$\underline{\underline{\lambda}}_d$ to consist of a uniform
expansion and a uniaxial, volume-preserving deformation 
along the pitch axis, $\underline{x}$:
\be
\underline{\underline{\lambda}}_d= \Phi^{-1/3} \;\;
\left(\lambda^{-1/2} \;\underline{\underline{I}}\;+\;
(\lambda-\lambda^{-1/2})\;\underline{x}\;\underline{x} \right),
\ee 
reflecting that, on coarse graining, the cholesteric is
uniaxial about the pitch axis. $\Phi$ is the volume fraction of
network after swelling. The volume $V_o$ of the network increases
to $V_o/\Phi$ with the additional solvent.

Substituting (2) and (3) into (1) leads to:
\be
{2\Phi^{2/3}\;F_n \over k_BT}
={1 \over \lambda} {(r-1)^2 \over r} \sin ^2 (\theta - q_o x)
+\lambda^2+{2 \over \lambda} \; .\label{1stF} \ee 
The $\sin^2$ term gives the director anchoring to the network.  
For small
rotations, $\omega$, relative to the matrix it gives the $D_1
\omega^2$ term of  the de Gennes continuum mechanics approach
\cite{dg1,olm}.  Note that without network anchoring, $\partial
F_n/\partial \lambda$ leads to $\lambda=1$ and there is no
tendency to relax shape (at a given dilation). From the $\sin^2$
term, we deduce that the anchoring coefficient after swelling,
$\tilde D_1$, is related to that before swelling $D_1$:
\be
\tilde D_1
={\Phi^{1/3}\over \lambda}D_1; \quad \quad D_1={(r-1)^2 \over
r}Y_o; \ee with the chain density defined as $n_n={N \over V_o}$,
the network shear elastic constant being $ Y_o={n_n k_BT}$, and
$N$ being the total number of chains.

In the Frank energy, $F_f$, we retain only twist:
\be
{2 \Phi \over V_o  K_2 }F_f = (\underline{n} \cdot \nabla \times
\underline{n}+q)^2 \equiv (- d\theta /d x +q)^2 \; . \ee  For
simplicity we assume the local nematic order, and therefore $K_2$,
remain unchanged during swelling - for instance if the chiral
solvent has the same nematogenic properties as the polymer's own
nematogenic elements. The pitch wave vector $q$ will be different
from the originally imprinted $q_o$ due to deformation
${\Phi^{1/3} / \lambda}$ along the $x$ axis from swelling and
shape change. The solvent chirality, $q_s$, further
modifies $q$:
\be
q={\Phi^{1/3} \over \lambda}q_o- q_s(2\phi-1) \label{newq}\; . 
\ee
Thus the first term corresponds to geometry and the second to the
left-right imbalance of the solvent; the volume fractions of the
left and right handed solvents are $\phi_l=\phi$ and
$\phi_r=1-\phi$.  The difference from racemic, $\phi - 1/2$,
allows the twist, $\pm q_s$, of the enantiomers to express itself.

\begin{figure}[htb]
  \begin{center}
    \leavevmode
    \vspace*{0.2cm}
    \epsfysize=1.45in
    \epsfbox{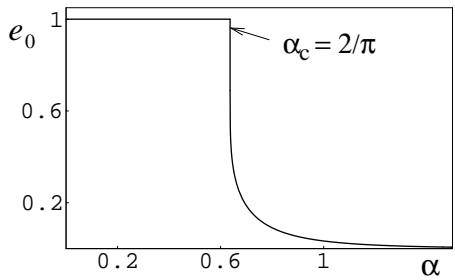}
    \caption{Imprinting efficiency vs.\ network parameter $\alpha$. 
The slope at the critical point $\alpha=\alpha_c^+$ diverges.}
  \label{efficient}
  \end{center}
\end{figure}

Imprinting is then a competition between the elastic anchoring
energy, $F_n$, which resists director rotation away from the
direction in which it was formed, and the Frank energy, $F_f$,
which drives $ d\theta /d x$ towards the current chiral twist $q$.
If $q$ is different from the value $q_o$ remembered by the
network, then twist towards this new value may occur depending
upon whether the anchoring, $\tilde D_1 q^2$ or the Frank, $K_2$
energy prevails.  This balance is quantified by $\tilde \alpha$,
the geometric ratio of the two energy scales: $$ \tilde
\alpha=\sqrt{K_2/\tilde D_1}\;q ={\sqrt{\lambda}\; q\over
\Phi^{1/6} q_o} \alpha_o; \quad \alpha_o=\sqrt{K_2/D_1}\; q_o \; .
$$  
In reference \cite{us1}, chiral solvent was replaced altogether, 
leaving a Frank
penalty $K_2 q_o^2$. Imprinting was successful for $\alpha_o <
2/\pi$ and the twists were retained.  Above $2/\pi$ one
must optimise over  $\theta$-profiles to find the minimal free
energy, and some untwisting occurs (the efficiency of 
imprinting is $< 1$, see figure \ref{efficient}).
\vspace*{-0.02in}
\noindent Next we compare $\tilde \alpha$ with $2/\pi$ and examine 
the preferential absorption of one handedness over the other.

The mixing free energy density can be written as: \ber f_m&=&{n_m}
k_BT [ \chi_{lr} \phi (1-\phi) (1-\Phi)^2+ \phi(1-\Phi)\ln \phi+
\nonumber \\ & &(1-\phi)(1-\Phi)\ln(1-\phi)+(1-\Phi)\ln(1-\Phi)+
\nonumber \\ &
&\chi_{ln}\phi(1-\Phi)\Phi+\chi_{rn}(1-\phi)(1-\Phi)\Phi ] \eer
where $n_m = 1/v_m$ is the solvent number density (with $v_m$ the
volume occupied by a single solvent molecule). The usual $\chi$
interaction parameters describe the interaction of left and right
molecules with each other ($lr$) and with the network ($ln$ and
$rn$). The latter are equal by symmetry if the network polymers
are achiral.
 The free energy of the entire system is:
\be
F=N\;F_n+F_f+{ V_o \over \Phi}\;f_m. \ee \noindent As in the case
of imprinting, we need to minimise energy over the profile of
nematic director orientation $\theta$ as we go down the pitch
axis.  We can do this first since the mixing part does not contain
$\theta$, and then we need to find the optimal $\lambda$.  We do
this separately for different regimes of imprinting parameter
$\tilde \alpha$.

\vspace{.25cm} \noindent {\it The small $\tilde \alpha$ limit.} In
this regime of high imprinting efficiency, the director
distribution is preserved. Thus there is no $D_1$ penalty, but a
Frank penalty for not having the currently desired twist $q$: $$
f(\phi,\lambda)=F \Phi/V_o={K_2 \over 2}q^2 + {\Phi^{1/3} Y_o
\over 2} \; (\;\lambda^2+{2 \over \lambda})+f_m $$ where $f_m=F_m
\Phi/V_o$, and  $q$  depends on solvent composition, see eq.
(\ref{newq}). Minimising $f$ over $\lambda$ gives a 4${\rm ^{th}}$
order polynomial equation in $\lambda$ (given the $\lambda$
dependence of $q$):
\be
-K_2{q q_o \over \lambda^2} + {Y_o}(\lambda-{1\over \lambda^2})=0
\; .\ee However,  assuming the uniaxial relaxation $\epsilon
=\lambda -1$ is small, one can easily solve the above:
\be
\epsilon \approx -{K_2\over Y_o} \;(\phi^{1/3}q_o-q_s(2\phi-1))q_o
\; . \ee Typically, $K_2\sim 10^{-11}{\rm N}$, $Y_o \sim 10^5 {\rm
Pa}$ and $q_{o,s}\sim 10^5 {\rm m}^{-1}$  and it appears that the
assumption of small $\epsilon$ ($\sim 10^{-3}$) applies to most
materials. Note that the explicit $\Phi$ dependence has vanished
in this limit, and the shape relaxation is independent of the
degree of swelling except indirectly through $\phi$. Substituting
$\lambda\approx 1$ back, we have:
\be
f(\phi)={K_2 \over 2}\tilde q^2 + {3\Phi^{1/3}\over 2}Y_o+f_m \ee
where $\tilde q={\Phi^{1/3}}q_o - q_s(2\phi-1)$. The chemical
potential difference $\Delta \mu =\mu_l - \mu_r = ({1 / [n_m
(1-\Phi)]})\; {\partial f / \partial \phi}$ is: \ber \frac{\Delta
\mu} {k_BT} = \eta +\chi_{lr}(1-2\phi)(1-\Phi) +\ln {\phi \over
1-\phi}  \nonumber \eer where $ \eta={-2K_2\; \tilde q q_s} / [
n_mk_BT(1-\Phi)]$.
In the reservoir (no network, $\Phi \rightarrow 0$,
$\phi_l \rightarrow \phi_o$ and $\phi_r \rightarrow 1-\phi_o$), 
the mixing energy density
$f_{r}$ is: 
$$ 
{f_{r} \over n_mk_BT}= \chi_{lr} \phi_o (1-\phi_o)
+ \phi_o \ln\phi_o+(1-\phi_o)\ln(1-\phi_o) 
$$ 
which yields the
chemical potential difference: 
$$
\frac{\Delta \mu_{r}} {k_BT} =
{1\over n_m {k_BT} } {\partial f_{r} \over \partial \phi_o} =
\chi_{lr}(1-2 \phi_o) + \ln {\phi_o \over 1-\phi_o}\; . 
$$ 
We now equate this difference to that of the network: \ber &
&\chi_{lr}(1-2 \phi_o) + \ln {\phi_o \over 1-\phi_o} = \nonumber
\\ & &\eta +\chi_{lr}(1-2\phi)(1-\Phi) +\ln {\phi \over 1-\phi} \;
. 
\eer
Suppose the $\chi_{lr}$ terms are negligible and that $K_2\;
\tilde q q_s \ll n_mk_BT$, {\it i.e.} $\eta$ is small compared 
to $1$. We then have for the l-concentration,
$\phi$, in the network, given $\phi_o$ in the reservoir:
\be
\phi \approx {1-\eta \over 1-\eta \phi_o}\;\phi_o \ee We should also
equate the osmotic pressures ($\partial F/\partial \Phi$) to
determine the equilibrium values of $\Phi$, in addition to the
$\phi$ obtained above. In practice, $\Phi$ is readily fixed
experimentally; we take it as a known system parameter.

\vspace{.25cm} \noindent {\it The large $\tilde \alpha$ limit.} In
this regime ($\tilde \alpha \gg 2/\pi$) of low imprinting
efficiency, the director unwinds to the currently desired twist
$q$. There is then no Frank penalty but  a $D_1$ penalty  since
the anchoring has been violated:  $$ f(\phi,\lambda)=F
\Phi/V_o={\tilde D_1 \over 4} + {\Phi^{1/3}\over 2 } Y_o \;
(\;\lambda^2+{2 \over \lambda})+f_m \; .$$  Minimisation over
$\lambda$ yields $\tilde \lambda = \left[{(r+1)^2 / 4r} \right]^{1/3}$
and  the free energy: \be f(\phi)={3 \Phi^{1/3} Y_o\over
2}\left[{(r+1)^2 \over 4r}\right]^{2/3} +f_m \; ,\ee from which it
follows that: \ber \frac{\Delta \mu} {k_BT} =
\chi_{lr}(1-2\phi)(1-\Phi) +\ln {\phi \over 1-\phi} \nonumber\; .
\eer Equating the chemical potential difference to $\Delta \mu_r$
leads to the equilibrium value of $\phi$. If we assume that the
$\chi_{lr}$ term is small so that $\chi_{lr}^2$ and higher order
terms in $\chi_{lr}$ can be ignored, then the network $\phi$, is
close to that, $\phi_o$, in the reservoir:
\be
\phi/\phi_o \approx 1+\chi_{lr}(1-2\phi_o)(1-\phi_o)\Phi \; .\ee

\begin{figure}[htb]
\begin{center}
\epsfysize=1.8in \epsfbox{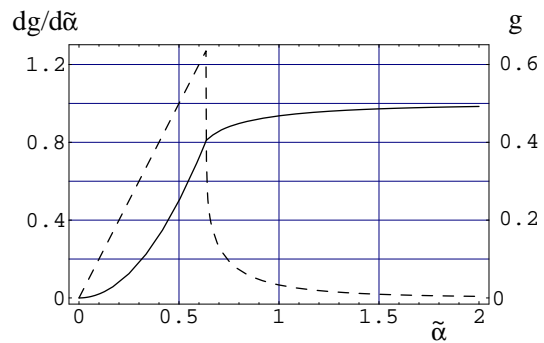}
\vspace*{-0.1in}
\end{center}
 \caption{
Director profile free energy $g(\tilde \alpha)$ and derivative
$dg(\tilde \alpha)/d\tilde \alpha$ (dashed line) against imprinting power
$\tilde \alpha$.} \label{gplots}
\end{figure}

\begin{figure}[htb]
\begin{center}
\epsfysize=1.8in \epsfbox{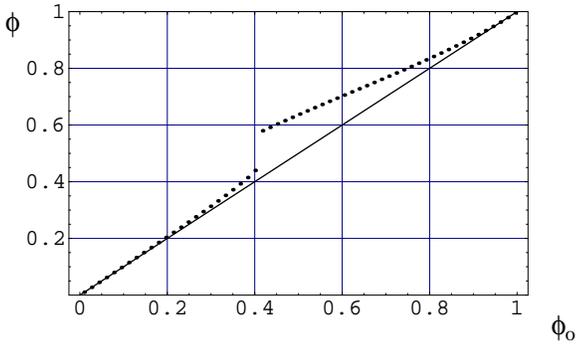}
\end{center}
 \caption{
Chiral demixing in the network, $\phi$, against reservoir solvent
handedness, $\phi_o$. See text for parameters used.} \label{sols}
\end{figure}

\vspace{.25cm} \noindent{\it The intermediate $\tilde \alpha$
regime.} In general, minimisation over the $\theta$-profile
\cite{us1} gives:
\be
f(\phi,\lambda)=g(\tilde \alpha) \tilde D_1/2+ {\Phi^{1/3}\over 2
} Y_o \; (\;\lambda^2+{2 \over \lambda})+f_m 
\ee 
with the profile
free energy  $g(\tilde \alpha)$  given by: 
\ber 
g(\tilde \alpha)&=&\tilde \alpha^2 \quad \quad \quad \quad  \quad
\quad \;\; {\rm for} \quad \tilde \alpha < 2/\pi \nonumber
\\ &=& \tilde \alpha^2 - k^{-2} +1 \quad \quad {\rm for} \quad \tilde
\alpha > 2/\pi 
\eer 
and where $k$ is related to $\tilde \alpha$
via $ \tilde \alpha = {2 {\cal E}(k) / \pi k}$, (see
Fig.~\ref{gplots} for $g$ and its derivative).  ${\cal E}$ is the
complete elliptic integral of the second kind. Minimisation of
$f(\phi,\lambda)$ over $\lambda$ gives a condition for the optimum
$\lambda=\tilde \lambda$:
\be
\left[ {\partial g \over \partial \tilde \alpha}\; {\partial
\tilde \alpha \over \partial \lambda} -{g\over \lambda}\;\right]\;
{(r-1)^2 \over 2\;\lambda\; r\;} + (\;\lambda-{1 \over \lambda^2})
=0 \ee Equating the network and reservoir exchange chemical
potentials yields:
\be
-\;\gamma\; {\partial g \over \partial \tilde \alpha}+ {1 \over
n_m k_BT}\; {1\over 1-\Phi}\;{\partial f_m \over \partial \phi}
={1 \over n_m k_BT} \;{\partial f_{r} \over \partial \phi_o}
\label{maineq} \ee where the prefactor $\gamma$ is given by
$$\gamma \; = \;{(r-1)^2 \over r} \;{n_n \over n_m}\; {\Phi^{1/6}
\over 1- \Phi }\;{q_s \over q_o} \; {\alpha_o \over \tilde
\lambda^{1/2}} .$$ Substituting the expressions for $f_{m}$ and
$f_{r}$ leads to \ber -\;\gamma\; {\partial g \over \partial
\tilde \alpha}&+& \chi_{lr}(1-2\phi)(1-\Phi) +\ln {\phi \over
1-\phi} \nonumber \\ &=&\chi_{lr}(1-2\phi_o) +\ln {\phi_o \over
1-\phi_o} \label{maineqsim} \eer The first term on the left hand
side of the equation is the contribution from the solvent-network
interaction, which discriminates between the solvent handednesses.
If $\gamma$ and $\Phi$ tend to zero, we recover the situation
where $\phi=\phi_o$.
\noindent Assuming that $\Phi$ and $\tilde \lambda$ do not depend
sensitively on $\tilde \alpha$ (which seems justified by the
independence of $\tilde \lambda$ on $\tilde \alpha$ in both the
small and large $\tilde \alpha$ limits), we can solve equation
(\ref{maineq}) by approximating $\tilde \lambda \approx 1$ and
taking a set of typical illustrative parameters:
$
\Phi =0.9, \;\; \alpha_o = {2 / \pi}, \;\; r = 2, \;\; {n_n / n_m}
= 0.2, \;\; {q_s / q_o} = 1, \;\; {\rm and} \;\; \chi_{lr}=0.
$
The solution is shown in figure \ref{sols}.
Fig.\ \ref{sols} shows that $\phi > \phi_o$ is generally
satisfied: the solvent handedness which agrees with the network is
favoured for absorption. One can utilize this to devise a
``pumping" cycle for extracting one component from a racemic
mixture. An elastomer, exposed to a given mixture, preferentially
absorbs one of the components. The solvent mixture, recovered from
stretching the swollen elastomer and thereby switching off its
chiral effect, can then be used in the next cycle of purification.

\begin{figure}[htb]
\begin{center}
\epsfysize=1.8in \epsfbox{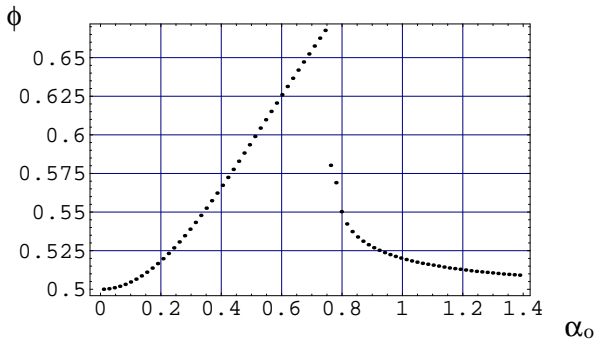}
\end{center}
 \caption{
Demixing, $\phi$ versus formation imprinting power, $\alpha_o$,
with constant reservoir concentration $\phi_o=0.5$.}
\label{alpha0}
\end{figure}

If we choose to fix the reservoir concentration $\phi_o=0.5$ and
vary the network's imprinting power at formation, $\alpha_o$, we
can see from figure \ref{alpha0} that the maximal resolving power
is obtained when the imprinting power of the resulting network is
near the transition point $\alpha=2/\pi$.

\begin{figure}[htb]
\begin{center}
\epsfysize=1.8in \epsfbox{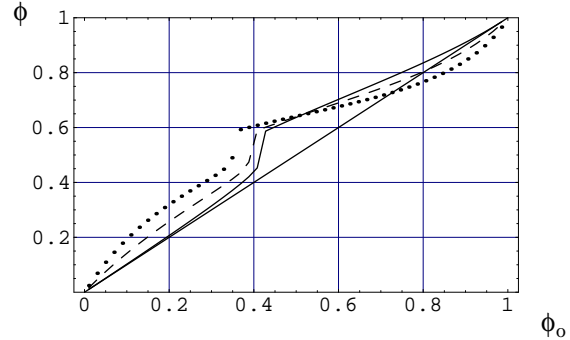}
\end{center}
\vspace*{-0.25cm}
 \caption{
Demixing dependence on $\chi_{lr}$, $\phi$ versus reservoir $\phi_o$ for
$\chi_{lr}=0$, $0.5$ (dashed), $1$ (dotted).} \label{chi}
\end{figure}

Finally, we examine the effect of $\chi_{lr}$. It has been
hitherto set to zero, that is we assume that any spontaneous
tendency for the molecules of opposite handedness to demix is
small. We see in figure \ref{chi} that a positive value of
$\chi_{lr}$ favours demixing at small $\phi_o$. Repeated cycles of
our chiral pump will push $\phi$ to the intersection of the curve
with the line $\phi=\phi_o$.

We have shown that the  chiral structures offered by imprinting
allows one to  preferentially absorb one handedness from a racemic
mixture into a network.  Moreover, since the chirality is
mechanically tuneable and there is no material of intrinsic
chirality in the network, one can then  release by mechanical
fields the absorbed solvent in which a chiral imbalance has been
achieved.


We thank E.~M.~Terentjev and M.~E.\ Cates for useful discussions.
YM is grateful to Christ's College, Cambridge for a Dow Senior
Research Fellowship.


\end{multicols}
\end{document}